\documentclass[runningheads,a4paper]{llncs}

\usepackage{amssymb}
\usepackage{graphicx}
\usepackage{times}
\usepackage{epsfig}
\usepackage{amsmath}
\usepackage{amssymb}
 \usepackage{booktabs}
 \usepackage{multirow}
\usepackage{pifont}
\usepackage{caption}
\usepackage{subcaption}
\usepackage{xcolor}
\usepackage{float}
\usepackage{url}
\urldef{\mailsa}\path|{alfred.hofmann, ursula.barth, ingrid.haas, frank.holzwarth,|
\urldef{\mailsb}\path|anna.kramer, leonie.kunz, christine.reiss, nicole.sator,|
\urldef{\mailsc}\path|erika.siebert-cole, peter.strasser, lncs}@springer.com|    
\newcommand{\keywords}[1]{\par\addvspace\baselineskip
\noindent\keywordname\enspace\ignorespaces#1}
\usepackage[backref=page]{hyperref}
\begin{document}

\mainmatter  

\title{Impact of loss function in Deep Learning methods for accurate retinal vessel segmentation}


%
%
\author{Daniela Herrera\inst{1}, Gilberto Ochoa-Ruiz\inst{1}, Miguel Gonzalez-Mendoza\inst{1} and \\
Christian Mata\inst{2,3} }
\authorrunning{Lecture Notes in Computer Science: Authors' Instructions}

\institute{Tecnologico de Monterrey, School of Engineering and Sciences, Mexico. \and
Pediatric Computational Imaging Research Group, Hospital Sant Joan de Déu
\and Research Centre for Biomedical Engineering (CREB), Barcelona East School of Engineering, Universitat Politècnica de Catalunya, 08019 Barcelona, Spain}


%
%

\toctitle{Retinal vessel Deep Learning segmentation comparison}
\tocauthor{Authors' Instructions}
\maketitle

\begin{abstract}
The retinal vessel network studied through fundus images contributes to the diagnosis of multiple diseases not only found in the eye. The segmentation of this system may help the specialized task of analyzing these images by assisting in the quantification of morphological characteristics. Due to its relevance, several Deep Learning-based architectures have been tested for tackling this problem automatically. However, the impact of loss function selection on the segmentation of the intricate retinal blood vessel system hasn't been systematically evaluated. In this work, we present the comparison of the loss functions Binary Cross Entropy, Dice, Tversky, and Combo loss using the deep learning architectures (i.e. U-Net, Attention U-Net, and Nested UNet) with the DRIVE dataset. Their performance is assessed using four metrics: the AUC, the mean squared error, the dice score, and the Hausdorff distance. The models were trained with the same number of parameters and epochs. Using dice score and AUC, the best combination was SA-UNet with Combo loss, which had an average of 0.9442 and 0.809 respectively. The best average of Hausdorff distance and mean square error were obtained using the Nested U-Net with the Dice loss function, which had an average of 6.32 and 0.0241 respectively. The results showed that there is a significant difference in the selection of loss function.  

\end{abstract}

\section{Introduction}
\label{introduction}

Analyzing eye fundus images is relevant for the identification of not only eye diseases but also systemic diseases, since the retina is susceptible to changes in the blood circulation in the brain  \cite{Abramoff_2010}. Usually, the study of retinal vessel structure is conducted through non-invasive techniques. In particular, the characterization and segmentation of retinal images is relevant for evaluating and assisting in the identification of cardiovascular diseases,  hypertension, strokes, and retinopathies \cite{Miri_2017}. Diabetic retinopathy is present in 80\% to 85\% of the patients who have diabetes for more than 10 years \cite{Sisodia_2017} and the gold standard for detecting it is through fundus imaging \cite{Kumari_2021}.
 
 The examination of the images obtained by this means and the assessment of morphological changes in this structure is a specialized and operator's dependent task. As a consequence of the process of projecting the 3-D semi-transparent retinal tissue into a 2-D imaging plane \cite{Abramoff_2010}, the evaluation of the image faces various challenges. Quantification of the structure analysis of the vessels (i.e.,symmetry, width, length), to understand the pathological changes produced, is made through image processing and segmentation of the vessels \cite{SAUNetSA}. A manual segmentation of these vessels is a time consuming and subjective task and therefore tools for automating this process are necessary to aid in the diagnostic process. Even though the images obtained have a high resolution, the contrast of the background and the blood vessels are similar to each other, making the task of segmentation and identification difficult \cite{Duran_2021}. 
 \par
The segmentation of the retinal vascular network has been addressed using deep learning with several architectures. There are multiple works based on the U-Net structure: SA-UNet \cite{SAUNetSA}, Attention-UNet \cite{AttentionUnet}, Generative Adversarial Networks \cite{Son_19}, among others. Attention modules have been added to the principal architectures since this mechanism tells the optimization process in which features to put more focus during the CNN weight updates \cite{CBAM_2018}, which may help in the segmentation of intricate patterns such as the vessel network. 

Besides the variety of architectures that exist for segmentation, the learning algorithm is instigated by the loss function, which should be selected depending on the objective \cite{Jadon_2020}. For evaluating segmentation results, there also exist multiple metrics that need to be selected considering the purpose and sensitivity of each one. The impact in the segmentation and evaluation of the quality by choosing loss functions and metrics is a challenge in deep learning segmentation applications.

In this paper, we explore recent archuectural strides (i.e. U-Net, Attention U-Net, SA-UNet, and Nested U-Nets) for segmentation to evaluate their performance in the segmentation retinal irrigation structures. Although comparisons have been made between the performance of architectures in retinal vessel segmentation \cite{Khanal_2020,Son_19}, the effect of loss function selection and its impact in the evaluation metric hasn't been reported in the literature to best of our knowledge. 

\begin{figure}
    \centering
    \includegraphics[width = 1\textwidth
    ]{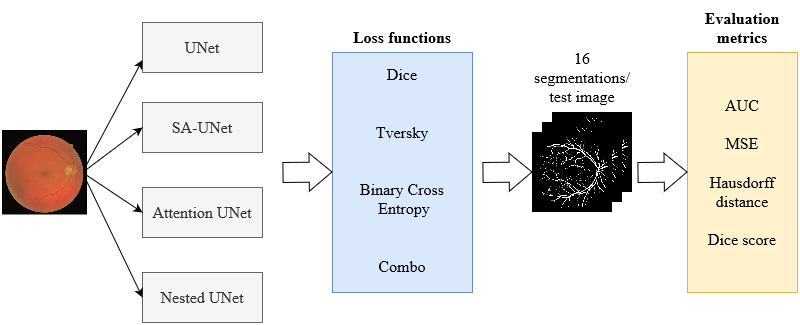}
    
    \caption{Pipeline for our experiments. Four models were trained using four loss functions, giving as a result 16 images for each test image. The results were assessed using four segmentation metrics.}
    \label{fig: pipeline}
\end{figure}

To conduct our experiments we used the Digital Retinal Images for Vessel Extraction (DRIVE) \cite{DRIVE} dataset and we explored several modern deep learning-based segmentation methods. The four deep learning-based models were trained in similar conditions with four loss functions, four metrics were used for assessing the results. The pipeline of the experiments is observed in figure \ref{fig: pipeline}. In this study, we compare the loss function using different metrics with four models that are U-Net structure-based, some of them also present attention modules.

The rest of this paper is organized as follows. In section \ref{State_of_the_art}, we discuss previous studies related to segmentation and specialized works in retinal vessel segmentation. Section \ref{data_methods} contains the description of the dataset, loss functions, metrics, and experimentation details. The results are presented in section \ref{results}. The conclusion and future work are presented in section \ref{conclusions_future_work}.

 \vspace{-0.5cm}

\section{State of the art}
\label{State_of_the_art}
The segmentation of retinal vessel structures aims to identify and distinguish the retina's vasculature from other anatomical structures that conform the background of the image \cite{Review_retina}. Although there have been improvements and developments in these tasks in recent years, robust segmentation still faces challenges such as the small number of available datasets for training, most of them having 50 images at most. The misclassification of vessels as background is related to the high imbalance problem found in these images, since typically only 10\% are classified as vessel \cite{Khanal_2020}.

The segmentation methods that have been used for this type of images can be grouped into three groups, based on the on the classification did by Abdulsahib et al. \cite{Review_retina}. In the first group, there are the rule-based methods, which were the first approaches used for this problem. These methods are fully unsupervised, meaning that they don't require any annotations. The second group includes machine learning (ML) techniques, which are usually supervised methods; the  feature extraction of the fundus image is done manually. Lastly, we consider separately a sub group from ML techniques, which are the Deep Leaning methods, due to their large impact and variety of architectures for segmentation. In such approaches, the biggest advantage is that the features from the images are automatically extracted and combined \cite{Retinal_seg_DL}. This last group is the focus of this study. 

\subsection{Rule-based methods}
A filtering approach is the kernel-based methods, which are constructed on the basis of the retinal vessels intensity distribution. These methods are capable of detecting the boundaries of vessels \cite{Review_retina}. There are examples of this method applied to retinal vessel segmentation \cite{kernel_2}. The vessel tracking method uses seed points to detect vessel ridges. However, such methods require a high level of pre-processing to enhance the sizes and orientation of the vessels. Another conventionally used method are the mathematical morphological operations, which are associated with the shapes found in the features of an image instead of the intensities in the pixels \cite{Review_retina}. The advantage of these methods is that they don't require any form of labeling. However, their performance does not usually surpass the supervised methods \cite{Retinal_seg_DL}.

\subsection{Machine Learning methods}
 ML-based methods require a manual extraction of local descriptors from the images that are then followed by a classifier. They have shown a better performance than the conventional methods previously mentioned \cite{State_minimalist}. An example of these algorithms is the k-nearest neighbors classifier (KNN) where the features obtained from the DRIVE dataset are classified \cite{DRIVE}. There are also examples of super vector machine (SVM) classifiers, an example is a semi-supervised approach using fully and weakly labeled \cite{SVM}. The limitations of these methods are that due to the lack of automatic feature extraction,  they lack the capability of generalization \cite{Retinal_seg_DL}. 
\subsection{Deep Learning methods}
Since the emergence of Convolutional Neural Networks (CNN) multiple architectures have been developed and applied to segmentation and other tasks related to computer vision. Their success in outperforming previous methods stems from the capability of such architectures to automatically learn features from raw data \cite{State_minimalist}.

There exist multiple types of CNN architectures. In this work, we will focus on a deep learning architectures created specifically for medical image segmentation, as it is the case of U-Net \cite{UNet} and its variants. The architecture has convolutional layers and is formed by a encoder that does the down sampling of the image using a max pooling. Then, the feature maps are up sampled in the decoder. Both stages of the architecture are communicated through skip connections to solve the degradation problem in deep neural networks. This architecture outperformed previous methods for multiple types of medical images. Due to its relevance in segmentation, this model is the base for the deep learning models selected in this study.

From the U-Net architecture there multiple variants derived from it. An example is the Nested U-Net  or U-Net ++ \cite{UNet++}. This model is also formed by a encoder and a decoder network. The difference is that the skip connections are formed of a series of nested dense convolutional blocks. The number of convolutional layers in the skip connection depends on the pyramid levels. \par
Variations in the segmentation architectures have been created by incorporating attention modules \cite{SAUNetSA,AttentionUnet,CBAM_2018}. These modules help focus on the important features and ignore the rest by improving the representation of interest and helping where to center \cite{CBAM_2018}. An example of the integration of an attention gate is the Attention U-Net \cite{AttentionUnet}. The module is located at each level of the skip connections, filtering the features and allowing the coefficients to be local, improving the performance against the global gating. Another example of attention modules applied to a U-Net is the Spatial Attention U-Net (SA-UNet) \cite{SAUNetSA}, which uses a spatial attention module located between the encoder a the decoder. Additionally this architecture adds a dropout convolutional blocks.

Although the segmentation of retinal vessels has improved through the development of deep learning based architectures, there is still a lack of comparative studies on how the decision of a metric, loss function, and model affects the segmentation. Therefore, in this work we compare them to understand how it affects the behavior of the segmented vessels, which can be thin,  intricate and with similar contrast as the background.


\section{Data and methods}
\label{data_methods}
The technical contribution of this work is the evaluation of the impact of four loss functions with four metrics on the retinal vessel segmentation using U-Net, SA-UNet, Attention U-Net, and Nested U-Net architectures. The experiments comprise a total of 16 segmentation comparisons. These experiments are summarized on figure \ref{fig: pipeline}. The methodology followed is detailed in the following section. 

\subsection{Dataset}
\label{Dataset}
\begin{figure}[htb]
    \centering 
    \begin{subfigure}{0.5\textwidth}
      \includegraphics[width=0.5\textwidth]{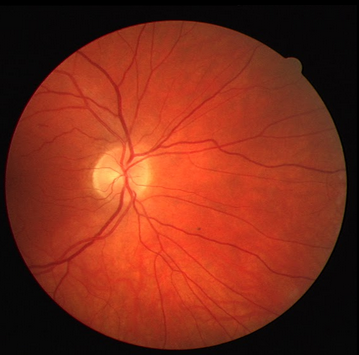}
      \centering
      \caption{Test image}
      \label{fig: DRIVE_Img}
    \end{subfigure}\hfil 
    \begin{subfigure}{0.5\textwidth}
      \includegraphics[width=.5\textwidth]{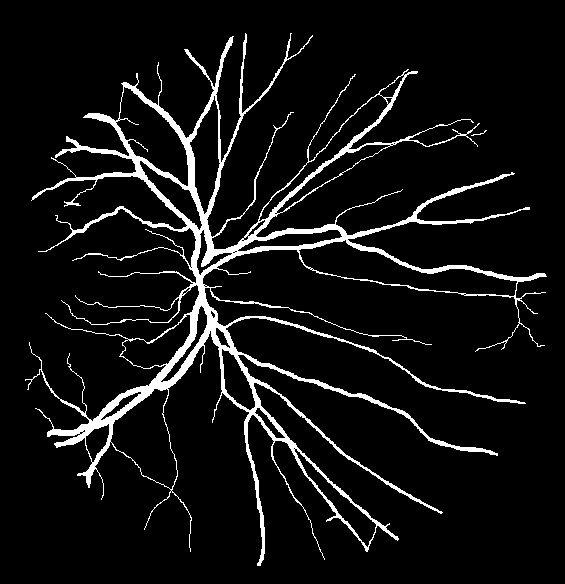}
      \centering
      \caption{Ground truth}
      \label{fig:DRIVE_label}
    \end{subfigure}\hfil 

\caption{Sample images for the DRIVE dataset \cite{DRIVE}.}
\label{fig:DRIVE_dataset}
\end{figure}

The images used in this work come from a retinal vessel segmentation dataset: the Digital Retinal Images for Vessel Extraction (DRIVE) \cite{DRIVE} dataset. It contains 40 images, of which 7 are abnormal pathology cases. We use the version with data augmentation that includes random rotation, Gaussian noise, color jittering, and flips (horizontal, vertical, and diagonal) from the SA-UNet paper \cite{SAUNetSA}. This increased the number of training images from 20 to 256 images. An example of the DRIVE dataset is shown in figure \ref{fig:DRIVE_dataset}; in the image there is the fundus image and the binary ground truth with the labels of the blood vessels.

\subsection{Segmentation metrics}
\label{Segmenatation metrics}
To evaluate the quality of the segmentation, there are numerous metrics that can be used depending on the data and the segmentation task \cite{Taha_metrics_2015}. For the first part of our experiments (loss function comparison), we used the of the following metrics: the Dice coefficient, the area under the ROC curve (AUC), means square error, and Hausdorff distance (HD).\par
The Dice coefficient computes the pair-wise overlap between the segmentation and ground truth divided by the common pixels between them: 

\begin{equation}
\text{DICE} = \frac{2|\text{Segmenation}\: \cap\: \text{Ground Truth}|}{|\text{Segmenation}| + |\text{Ground Truth}|}
\label{eqn: DICE}
\end{equation}

The Receiver Operating Characteristic (ROC) curve is the plot of the true positive rate (TPR) and the false positive rate (FPR). The area under the curve (AUC) was designed as a measure of accuracy. In the case of retinal vessel segmentation evaluates the accurately classified as background and vessels. The mean square error (MSE) averages the difference between the ground truth and the predicted pixel . The Hausdorff distance is a spatial-based metric measured in voxel size, which quantifies the distance between the ground truth and the segmentation. 

\subsection{Loss functions}
\label{loss_functions}
The loss functions help us in the mathematical representation of our segmentation objectives in deep learning to make it more accurate and faster \cite{Jadon_2020}. In this sense, it is expected that different loss functions will yield different segmentation results. Therefore, four loss functions were evaluated for this study: Dice, Tversky, Binary Cross Entropy, and  Combo loss. \\
The Dice loss function is based on the dice coefficient, explained in the previous section (equation \ref{eqn: DICE}), which minimizes the similarity between the ground truth and the segmentation results. The Tversky loss is based on the Dice loss and achieves a better balance between precision and recall, emphasizing the false negatives, by reshaping the dice loss \cite{Loss_odyssey_2021}. The Binary Cross Entropy loss measures the dissimilarity between two probability distributions. The Combo loss is a weighted summation between the Dice loss and a variation of the cross-entropy; this brings together the advantages from both losses \cite{Jadon_2020}.

\subsection{Deep Learning architectures}
\label{DL_architectures}

We tested four deep learning architectures for our experiments,  comparing various loss functions and multiple metrics, as described above.  The selected model are al based on the U-Net based architecture, originally proposed by Ronneberger et al. \cite{UNet}, this model is considered for the comparison. We selected the Spatial Attention U-Net (SA-UNet) proposed by Guo et al. \cite{SAUNetSA}. This architecture adds a spatial attention module between the encoder and decoder. It also integrates dropout convolutional blocks for reducing the overfitting 
 Attention UNet proposed by Oktay et al. \cite{AttentionUnet} was also selected; it adds an Attention Gate between the union of the skip connection and the decoder. The UNet++ is a nested UNet architecture where the encoder and decoder are connected through a series of nested dense convolutional blocks \cite{UNet++} is the last model selected.
\subsection{Training}

The four models were trained using equal parameters and circumstances. The SA-UNet model was trained using an implementation from the authors made in Keras \cite{SAUNetSA}. The Nested UNet implementation PyTorch made by its authors was used \cite{UNet++}, they also made an implementation for U-Net and Attention U-Net in PyTorch which was also used. The models were trained for 100 epochs, the weights that had the best validation score were the ones saved. The training was done using the Nvidia DGX workstation, two GPUs were used for each model except for Nested U-Nets were four GPUs were needed. The batch size was 4 and a learning rate was 0.001. 
\section{Results}
\label{results}
In previous works there were limitations presented in the segmentation of retinal blood vessels due to the class imbalance, similar contrast between vessels and background, and limited labeled datasets. The impact of loss function selection and metrics hasn't been compared using multiple models. Through out experiments, we expect to understand the influence of this selection on the quality of the segmentation of the vessel structure. \\
\begin{figure}[H]
    \centering
    \includegraphics[width = 1 \textwidth]{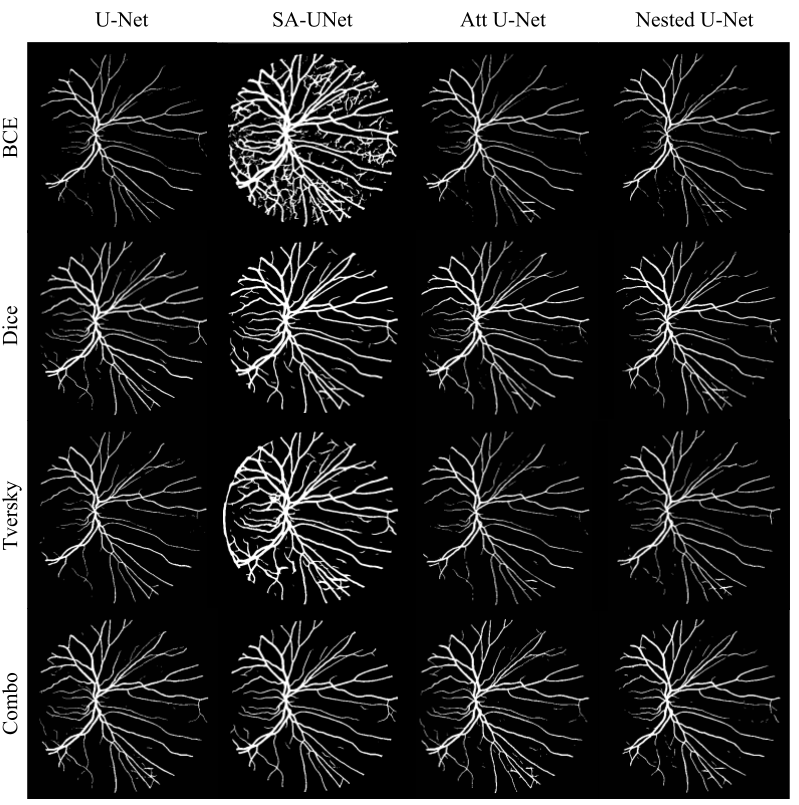}
    \caption{Visual comparison of the segmentation results obtained by the U-Net, Attention U-Net, SA-UNet and Nested UNet models using four loss functions (i.e. BCE, Dice, Tversky and Combo, respectively)}
    \label{fig: segmentations}
\end{figure}
The 16 segmentations results can be seen in figure \ref{fig: segmentations}. The figure shows the same test image for the multiple combinations. The original images showed in \ref{fig: DRIVE_Img} with its respective ground truth. The combination results are shown with  different loss functions (i.e. BCE, Dice, Tversky and Combo, respectively) and models (i.e. U-Net, SA-UNet, Attention U-Net, and Nested U-Net) respectively. 

 A summary of the quantitative results of the models, employing the four discussed loss functions and the performance of the model under the four discussed metrics of the experiments is shown on table \ref{tbl: loss funcitons}.  The table displays the average metric for each experiment. The symbol $\uparrow$ represents that the metric is better when is bigger, while the a $\downarrow$ symbol means that is better when is close to 0. The numbers in bold indicate the loss function that had the best performance for each metric. We can see that although they were trained for the same number of epochs, the metrics differ depending on the loss function. For the discussion that follows, the models are going to be first analyzed individually, then against each other and finally the loss functions performance is studied.\par
 
 \vspace{-0.5cm}

\renewcommand{\arraystretch}{1.5}
\begin{table}[H]
\caption{Results using multiple loss functions and metrics (average) for U-Net based architectures.}
\label{tbl: loss funcitons}
\centering
\begin{tabular}{llclllllclllllclllllcllll}
\hline
\multicolumn{1}{c|}{\textbf{Loss function}} & & \textbf{AUC $\uparrow$} & & & \textbf{MSE $\downarrow$} & & & \textbf{\begin{tabular}[c]{@{}c@{}}Hausdorff \\ distance $\downarrow$\end{tabular}} & & & \textbf{\begin{tabular}[c]{@{}l@{}}Dice \\ Score $\uparrow$\end{tabular}} &\multicolumn{1}{l|}{} & & \textbf{AUC $\uparrow$} & & & \textbf{MSE$\downarrow$} & & & \textbf{\begin{tabular}[c]{@{}c@{}}Hausdorff \\ distance $\downarrow$\end{tabular}} & & & \textbf{\begin{tabular}[c]{@{}l@{}}Dice \\ Score $\uparrow$\end{tabular}} &
   \\ \hline
 & \multicolumn{12}{c}{U-Net} & \multicolumn{12}{c}{SA-UNet} \\ \hline
\multicolumn{1}{c|}{Dice} & & \textbf{0.9387} & & & \textbf{0.0250} & & & \textbf{6.4304} & & & 0.7667 & \multicolumn{1}{l|}{}& & 0.9431 & & & 0.0565 & & & 9.3039 & & & 0.7327 &
   \\ 
\multicolumn{1}{c|}{Tversky} & & 0.9117 & & & 0.0260 & & & 6.5534 & & & 0.6776 & \multicolumn{1}{l|}{}& & \textbf{0.9442} & & & 0.0692 & & & 10.1913 & & & 0.6951 &
   \\ 
\multicolumn{1}{c|}{\begin{tabular}[c]{@{}c@{}}BCE\end{tabular}} & &
  0.8911 & & & 0.0286 & & & 6.7546 & & & 0.6276 & \multicolumn{1}{l|}{}& & 0.9197 & & & 0.1339 & & & 21.0970 & & & 0.5470 &
   \\ 
\multicolumn{1}{c|}{Combo} & & 0.9373 & & & 0.0259 & & & 6.4871 & & & 0.7668 &\multicolumn{1}{l|}{} & & 0.9335 & & & \textbf{0.0355} & & & \textbf{7.9129} & & & \textbf{0.809} &
   \\ \bottomrule
 & \multicolumn{12}{c}{Attention U-Net} & \multicolumn{12}{c}{Nested U-Net} \\ \hline
\multicolumn{1}{c|}{Dice} & & \multicolumn{1}{l}{\textbf{0.9392}} & & & \textbf{0.0251} & & & 6.4251 & & & 0.7695 & \multicolumn{1}{l|}{}& & 0.9318 & & & \textbf{0.0242} & & & \textbf{6.3211} & & & 0.7345 &
   \\ 
\multicolumn{1}{c|}{Tversky} & & 0.9094 & & & 0.0261 & & & 6.5723 & & & 0.6708 & \multicolumn{1}{l|}{}& & 0.8917 & & & 0.0279 & & & 6.7341 & & & 0.6434 &
   \\ 
\multicolumn{1}{c|}{\begin{tabular}[c]{@{}c@{}}BCE\end{tabular}} & &
  \multicolumn{1}{l}{0.9047} & & & 0.264 & & & 6.522 & & & 0.6470 & \multicolumn{1}{l|}{}& & 0.8863 & & & 0.0288 & & & 6.7185 & & & 0.6212 &
   \\ 
\multicolumn{1}{c|}{Combo} & & \multicolumn{1}{l}{0.9391} & & & 0.0259 & & & \textbf{6.4342} & & & \textbf{0.7731} & \multicolumn{1}{l|}{}& & \textbf{0.9344} & & & 0.0251 & & & 6.4174 & & & \textbf{0.7659} & \\\bottomrule
\end{tabular}
\end{table}

If we would base the performance on only one metric, we could ignore the overall performance. For the U-Net model, which is the base architecture, the best average score for AUC, MSE, and HD had it the dice loss function. For the case of the Dice Score the dice and combo loss had a similar performance with a difference of 0.0001. In this case, if we had only used the dice score for evaluating the results, we could have concluded that the best performance had it the combo loss (see Figure \ref{fig: segmentations}).  \\
For the SA-UNet in the case of AUC, the loss function that had the best performance was the Tversky loss. However, looking at the other metrics, the Combo loss had the best performance for the rest of the metrics of interest. 

For the segmentations obtained in this model, we can observe (figure \ref{fig: segmentations}) that the for the BCE loss there is the presence of noise and in the Tversky loss the limit of the fundus and the black background is also segmented. These details may be associated with a need of more epochs of training. 

\section{Discussion}
\label{discussion}

The performance of the Attention U-Net had the best results for the dice loss function using the AUC and MSE, while using the HD and dice score, the best one was performed by the combo loss. By observing the segmentations (figure \ref{fig: segmentations}) the one that segmented most fine vessels was the combo loss. \\
The Nested UNet presented the best performance with the dice loss function with the metrics that need to be minimized (HD and MSE). The combo loss obtained better results for the metrics that need to be maximized (AUC and dice score). This model was the one that required the most computational resources (4 GPUs instead of 2). 

\begin{figure}[]
    \centering
      \includegraphics[width=.99\linewidth]{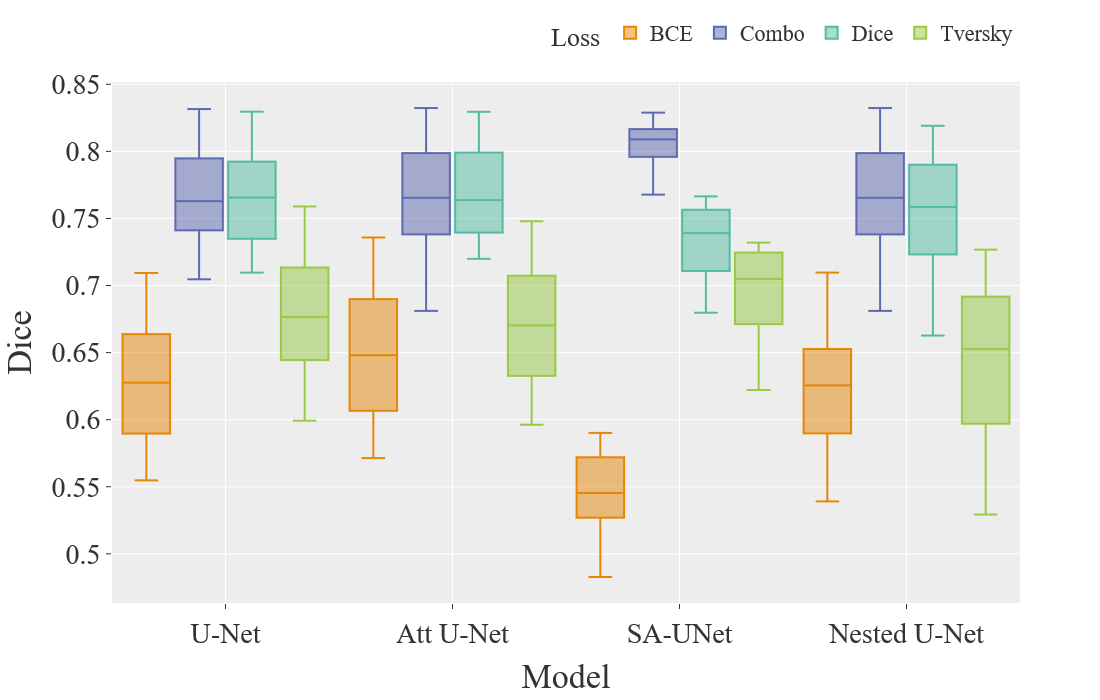}
    \caption{Results from Deep Leaning based architectures dice score comparison}
    \label{fig: boxplots}
\end{figure}

For comparing the segmentation results against models (UNet, SA-UNet, Attention UNet, and Nested UNets), the outcomes were evaluated using the Dice score $\uparrow$. The summary of the results can be seen in the boxplots from figure \ref{fig: boxplots}, where the distribution of the dice score for each loss function and model is shown. The SA-UNet is the one that had the best dice score results by using the combo loss since it had the highest average and the most compact results, meaning that they are less variable. \\
The model that had the worst performance using the dice score is also the SA-UNet, but using the BCE function. If we consider the average of the rest of the metrics in table \ref{tbl: loss funcitons} (AUC, MSE, and HD) we may have different results. The highest average for AUC was also obtained by the SA-UNet using the Tversky loss. The
smallest HD and MSE average was obtained by the Nested UNets using the dice loss. \par


The performance comparison by loss functions in the dice score, it is observed in the boxplot (figure \ref{fig: boxplots}) that the combo and dice loss results (dark and light blue) have the highest and most compact results, its average is above 0.7. The average of the dice score for the binary cross entropy loss (orange) was above 0.6, except in the case of the SA-UNet. For the case of the Tversky loss function, the average is also above 0.6 for the four models. Considering only averages form other metrics, the performance of the binary cross entropy loss, didn't have the highest average in any of the cases. \par
\begin{figure}[H]
    \centering
    \includegraphics[width = 0.8\textwidth]{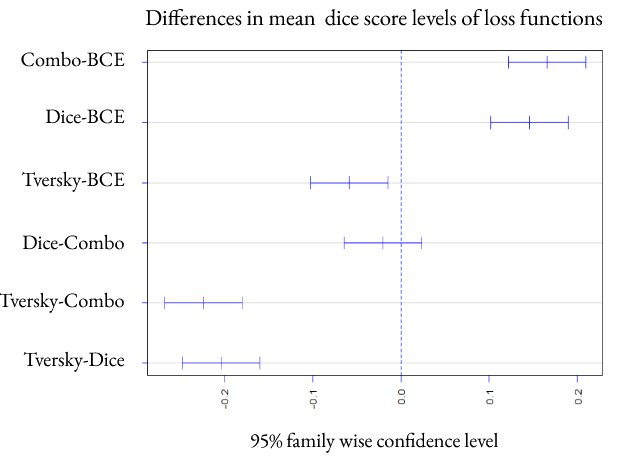}
     \caption{Results of Tukey post-hoc statistical tests for the segmentation results against dice score for each loss function.}
    \label{fig: tukey}
\end{figure}

To determine whether the results between each type of loss function were significantly different, a post-hoc Tukey test was performed using an alpha of 0.05, between the segmentation dice score results of the four architectures. The test does a pairwise mean comparison between the loss functions. The results are shown the in a graph (figure \ref{fig: tukey}). This test helps us determine whether the loss function selection indeed is significantly different.\\
The graph (figrue \ref{fig: tukey}) shows the dice score mean differences between the loss function and has a confidence interval of 95\%. The dotted line represent 0 difference between them. The loss functions that had a the greatest difference were the Tversky and the Combo loss. From the six comparisons, the only ones that were not significantly different, were the Dice and Combo loss function comparison. This may be associated with the fact that the Combo loss also uses the dice score in its formula.

\section{Conclusion and future work}
\label{conclusions_future_work}

 Considering the overall performance of the metrics, the best loss function was the Combo. The conclusion could have been different if it had only been determined by one metric. Therefore, the combination of the loss function and metric needs to be considered. For the Deep Leaning based architecture comparison, although they all were UNet based they had different results, which were also reflected in the statistical tests. For the Dice score, was the SA-Unet that had better performance, however using other metrics such as the mean square error, the best model was the Nested U-Net, this shows the sensitivity of the metrics and how it should be globally evaluated. \\
 With the epochs trained, each model had vessels that weren't correctly segmented, especially the most fine ones. This could be improved with more epochs of training for each model. There is still areas of opportunities for improving the segmentation. The feedback from an ophthalmologist can help in understanding the implications of these omissions. We observed that there was a significant impact on the selection of the loss function, this was reflected in the average of each metric, which was proved by a Tukey test.\\
For future work and a better understanding of the impact on retinal vessel segmentation, a comparison between other types of segmentation deep learning architectures should be done. The use of other retinal vessels segmentation datasets could also bring more comprehensive studies. In this study, it was shown that the selection of loss function has repercussions in the segmentation of retinal vessels, and for performing an intrinsic evaluation multiple metrics should be considered.


\section*{Acknowledgments}
\vspace{-0.1cm}
The authors wish to thank the AI Hub and the CIIOT at ITESM for their support for carrying the experiments reported in this paper in their NVIDIA's DGX computer.

\bibliographystyle{splncs04}
\bibliography{references}

\end{document}